\documentclass[]{spie}  

\usepackage{amsmath,amsfonts,amssymb}
\usepackage{graphicx}
\usepackage[colorlinks=true, allcolors=blue]{hyperref}

\newcommand{\arcsec}{$^{\prime\prime}$}
\newcommand{\arcmin}{$^{\prime}$}
\newcommand{\microm}{$\mu$m}

\title{The Planet as Exoplanet Analog Spectrograph (PEAS): Design and First-Light}

\author[a]{Emily C. Martin}
\author[a]{Andrew J. Skemer}
\author[b]{Matthew V. Radovan}
\author[b]{Steven L. Allen}
\author[b]{David Black}
\author[b]{William T. S. Deich}
\author[a]{Jonathan J. Fortney}
\author[b]{Gabriel Kruglikov}
\author[b]{Nicholas MacDonald}
\author[b]{David Marques}
\author[a]{Evan C. Morris}
\author[b]{Andrew C. Phillips}
\author[b]{Dale Sandford}
\author[c]{Julissa Villalobos Valencia}
\author[d]{Jason J. Wang}
\author[b]{Pavl Zachary}

\affil[a]{Department of Astronomy \& Astrophysics, University of California, Santa Cruz, 1156 High St. Santa Cruz, CA 95064}
\affil[b]{University of California Observatories, 1156 High St. Santa Cruz, CA 95064}
\affil[c]{Department of Mechanical and Aerospace Engineering, University of California, San Diego, 9500 Gilman Drive MC 0411
La Jolla, CA 92093-0411}
\affil[d]{Department of Astronomy, California Institute of Technology, 1200 E. California Blvd. Pasadena, CA 91125, USA}

\authorinfo{ Send correspondence to E.C.M.\\E-mail: emilymartin@ucsc.edu}

\begin{document} 
\maketitle

\begin{abstract}
Exoplanets are abundant in our galaxy and yet characterizing them remains a technical challenge. Solar System planets provide an opportunity to test the practical limitations of exoplanet observations with high signal-to-noise data that we cannot access for exoplanets. However, data on Solar System planets differ from exoplanets in that Solar System planets are spatially resolved while exoplanets are unresolved point-sources. We present a novel instrument designed to observe Solar System planets as though they are exoplanets, the Planet as Exoplanet Analog Spectrograph (PEAS). PEAS consists of a dedicated 0.5-m telescope and off-the-shelf optics, located at Lick Observatory. PEAS uses an integrating sphere to disk-integrate light from the Solar System planets, producing spatially mixed light more similar to the spectra we can obtain from exoplanets. This paper describes the general system design and early results of the PEAS instrument.
\end{abstract}

\keywords{Exoplanet instrumentation, Solar System instrumentation, off-the-shelf optics, spectrograph, Lick Observatory }

\section{INTRODUCTION}
\label{sec:intro}  

There are now thousands of confirmed exoplanets in our galaxy [\citenum{nexsci}] and we expect that on average, there is at least one exoplanet per star [\citenum{batalha2014}]. However, characterizing exoplanet atmospheres remains a significant technical challenge and it is possible that our interpretations of exoplanet atmospheres are systematically flawed. One method to validate exoplanet observational techniques is by applying those techniques to observations of Solar System planets, which have high signal-to-noise data that we cannot access for exoplanets. We have decades of ``ground-truth" data for Solar System planets from landers, rovers, probes, fly-by missions, and observations from both ground- and space-based telescopes. 

\begin{figure} [ht]
  \begin{center}
  \includegraphics[width=17cm]{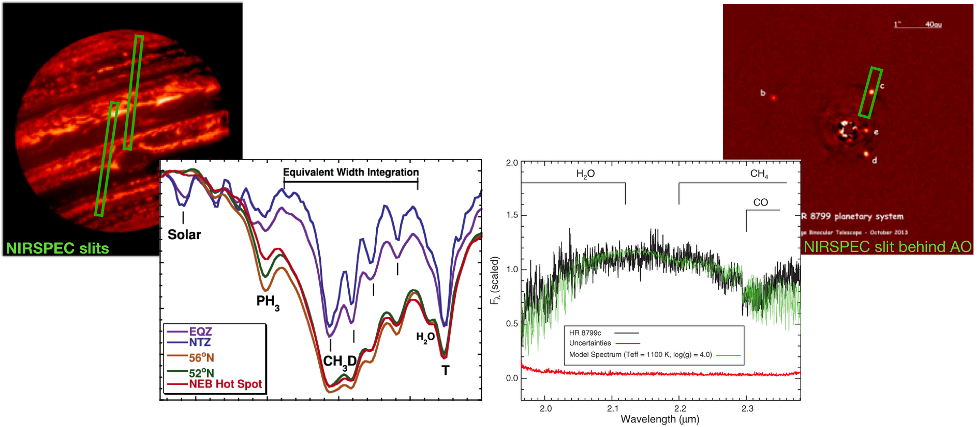}
  \end{center}
  \caption 
  { \label{fig:jupiterhr8799} 
\textit{Left}: 5 \microm \ image of Jupiter from IRTF with example NIRSPEC/Keck slits overlaid. The 4.6 \microm \ resolved spectra of Jupiter [\citenum{bjoraker2015}] vary widely depending on the location of the observed spectrum on the planet. \textit{Right}: HR8799 image [\citenum{maire2015}] and resolved OSIRIS/Keck spectrum of the full disk of HR8799c [\citenum{konopacky2013}]. The observed spectra of Jupiter are not comparable to the spectra of the directly observed HR 8799c planet because of the widely varying spatial scales. Jupiter's relative size on the sky is $\sim$ 40\arcsec \ while the HR8799 planets are single point sources. }
  \end{figure}

However, Solar System planet observations are not directly comparable to exoplanet observations because of an optics scaling problem. Solar System planets range in apparent angular size from several arc-seconds (Uranus, Neptune) to nearly an arc-minute (Jupiter, Venus), while every exoplanet we can resolve is a single point-source (Figure~\ref{fig:jupiterhr8799}). Historically, there have been several efforts to observe Solar System planets as point-source exoplanets, e.g., [\citenum{sagan1993}, \citenum{cowan2009}, \citenum{mayorga2016}]. Each of these studies used resolved photometry of Solar System planets summed up to a single pixel to produce ``disk-integrated" photometry. The most recent disk-integrated spectroscopy of the Jovian and Ice Giant planets are from [\citenum{karkoschka1994} and \citenum{Karkoschka1998}], both measured by using a long-slit spectrograph with a wide slit and observing while the planets transited the slit. Previously, there were no dedicated instruments for observing Solar System planets as if they were point-source exoplanets. Such a dedicated instrument would provide instantaneous measurements of Solar System planetary atmospheres and provide the potential for studying planetary variability on an on-going basis.

\begin{figure} [ht]
  \begin{center}
  \includegraphics[width=10cm]{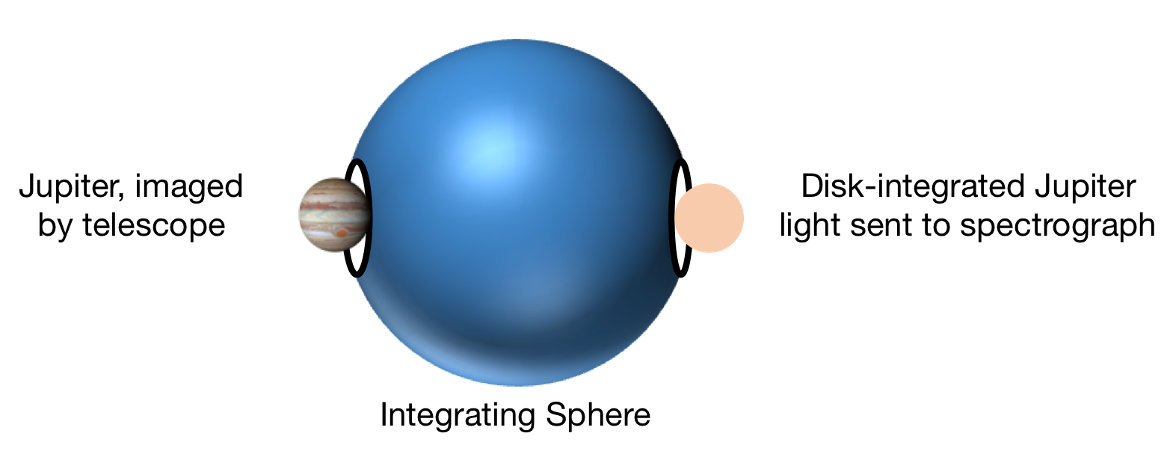}
  \end{center}
  \caption 
  { \label{fig:intsphere} 
An integrating sphere takes resolved images of planets (shown here, Jupiter) and disk-integrates the light, which is then fiber-fed to a spectrograph.}
  \end{figure}

We designed and built a new instrument for Lick Observatory called PEAS, the Planet as Exoplanet Analog Spectrograph, to observe disk-integrated spectra of Solar System planets as if they were point-source exoplanets. PEAS consists of a 0.51-m telescope from Planewave feeding an integrating sphere (Figure ~\ref{fig:intsphere}) which sends disk-integrated planet light via fiber to an  off-the-shelf spectrograph. An off-the-shelf imager provides simultaneous resolved imaging. The telescope and instrument are mounted on a rolling-pier mount from Planewave. PEAS is stored inside the Shane 3-m dome at Lick Observatory and rolls out onto the parking lot to observe.  Below, we describe the instrument design and present early results from first-light. 

\section{Instrument Design}
\begin{figure} [ht]
  \begin{center}
  \includegraphics[height=8.2cm]{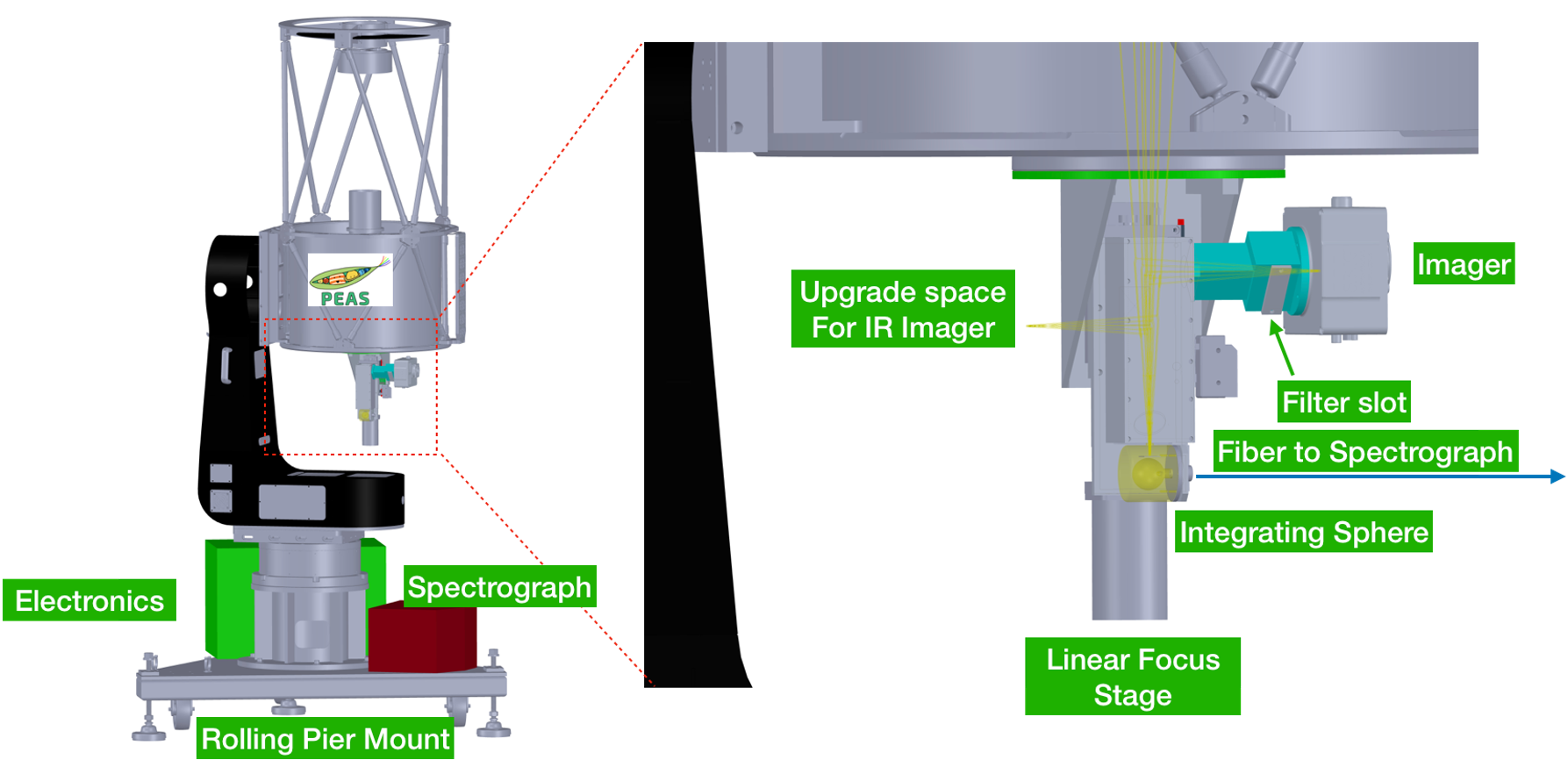}
  \end{center}
  \caption 
  { \label{fig:optomechoverview} 
\textit{Left}: SolidWorks Assembly drawing of the PEAS telescope and optics. The Planewave RC 20" (0.51-m) telescope attaches to the Planewave L500 mount which sits on the Planewave Rolling Pier Mount. The Andor spectrograph and electronics are also attached to the rolling pier. \textit{Right}: Expanded view of the optical path of light from the telescope in yellow rays. Light from the telescope passes through 4 sapphire windows acting as beamsplitters and compensators before focusing inside a 1" integrating sphere. Disk integrated light from the sphere is fed via fiber to the Andor Spectrograph. The first sapphire window sends reflected light into an SBIG imager and the second sapphire window has space for an eventual infrared imager upgrade. The back-end optics are fixed to a THK Linear Stage for focusing. }
  \end{figure}

PEAS has a simple optical and mechanical design using primarily off-the-shelf parts. An overview of the full system is shown in Figure \ref{fig:optomechoverview}. A 0.51-m telescope feeds an optical tube with windows that send light to an imager and an integrating sphere. A fiber output feeds light to an off-the-shelf spectrograph to observe disk-integrated spectra of Solar System planets.

\subsection{Optical Design}

PEAS uses a Planewave 20" (0.51-m) Ritchey-Chrétien (RC) telescope. The optical beam provided is f/7 with a 3544mm uncorrected focal length. The accompanying L500 Alt-Az direct-drive mount has $<$10 arcsec RMS pointing accuracy and 2 arcsec pointing accuracy while tracking sidereally. Tracking accuracy is $<$0.3 arcsec over 5 minutes. 

 \begin{figure} [ht]
  \begin{center}
  \includegraphics[width=15cm]{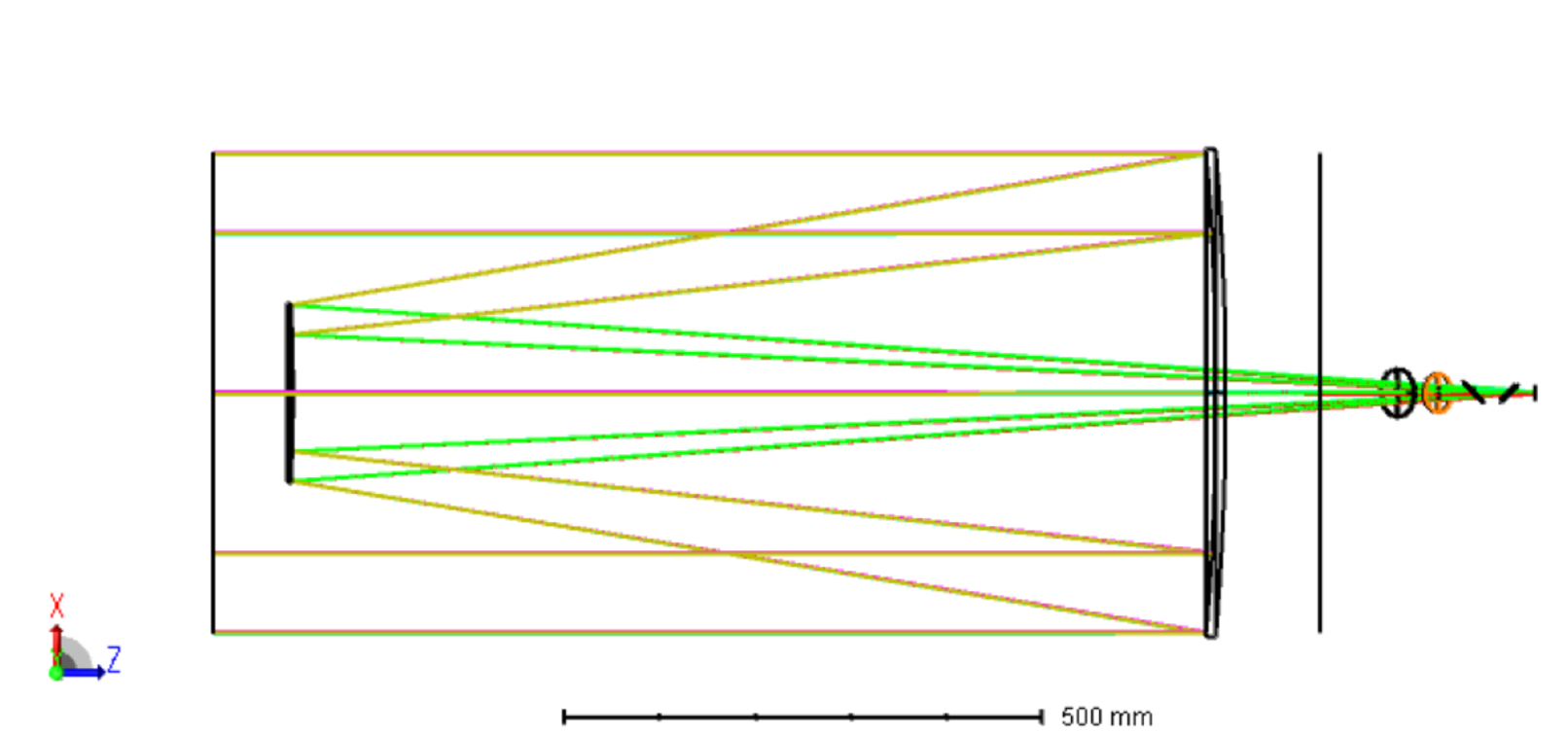}
  \end{center}
  \caption 
  { \label{fig:zemax} 
Zemax layout of the telescope and back-end optics. Four sapphire windows tilted at 45 degree angles along $+X$, $-X$, $+Y$ and $-Y$ axes act as beamsplitters and image compensators. Light comes to a focus inside the integrating sphere. Light reflecting off the first two windows is directed into the optical and infrared imaging channels (Configurations not shown).}
  \end{figure}

\begin{figure} [h!]
  \begin{center}
  \includegraphics[width=12cm]{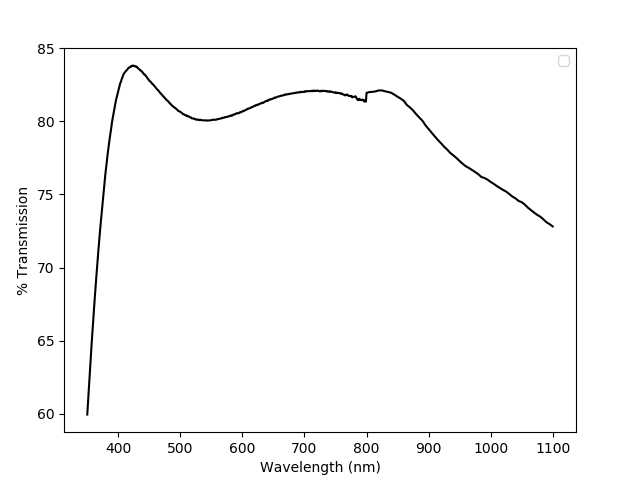}
  \end{center}
  \caption 
  { \label{fig:transmission} 
Transmission curve through all four sapphire windows, corrected for 45 degree angle of incidence. The 50mm and 40mm windows were only coated on one side to allow for higher reflection off the first surface into the optical and infrared imaging channels. The 30mm and 25mm windows were coated on both sides (total of 6 out of 8 sides coated). The abrupt jump at $\sim$800 nm is due to an artifact from the Cary Spectrophotometer transitioning to measure transmission from the optical to the near infrared.}
  \end{figure}

\begin{figure} [ht]
  \begin{center}
  \includegraphics[width=17cm]{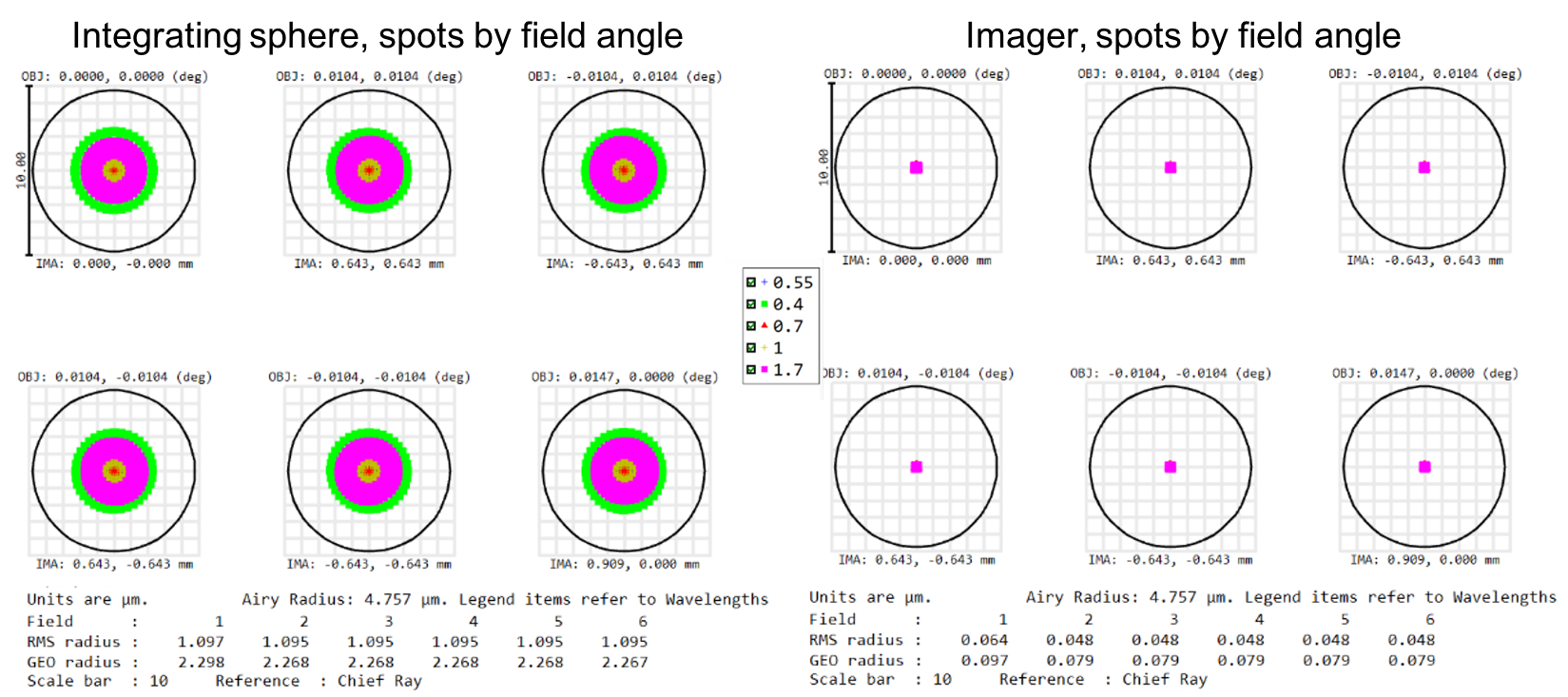}
  \end{center}
  \caption 
  { \label{fig:spots} 
Zemax spot diagrams for the integrating sphere (Left) and imager (Right) channels. Spots are diffraction limited at all field angles and wavelengths. Spots are colored by wavelength, as noted in the legend in the center.}
  \end{figure}

On the back end of the telescope, there is a series of 4 sapphire windows (Figure~\ref{fig:zemax}) tilted at 45 degree angles along $+X$, $-X$, $+Y$, and $-Y$. The first window acts as a beamsplitter to send reflected light to an optical imager for simultaneous imaging and tracking. The second window corrects geometric aberrations from the first window and will eventually send light to the infrared imaging channel in a future upgrade. The third and fourth windows correct for chromatic aberrations from the first two windows. The windows are 50 mm, 40 mm, 30 mm, and 25 mm in diameter, 2mm thick, random-cut from Knight Optical. They were coated in-house at UC Observatories with a custom designed anti-reflective (AR) coating (Figure~\ref{fig:transmission}). The largest windows, which serve as beamsplitters for the imagers, are only AR coated on the back sides. The two smallest windows have AR coatings on both sides. Following the windows, light is focused into an entrance hole on the integrating sphere. Figure~\ref{fig:spots} show the image quality at the image plane of the optical imager and the integrating sphere.

The integrating sphere is a 1" sphere from Spectral Products, coated with Spectralon. We use the 3.2 mm diameter SMA port as our entrance hole. The entrance hole has a depth of 9.85 mm, allowing a field of view within the integrating sphere of $\sim$ 150\arcsec. The output port is 8mm in diameter. We attached a custom fiber feed to hold the fiber bundle inside the exit port. The fiber bundle is a 91 fiber integral-field unit ferrule from the MaNGA/SDSS project [\citenum{drory2015}]. Each fiber has a 120 \microm \ core and 150 \microm \ outer diameter and NA of 0.22. At the other end of the bundle, the fibers are re-arranged into pseudo-slits held in V-groove blocks. We feed one pseudo slit of 30 science fibers + 2 sky fibers into our off-the-shelf spectrograph. 

The spectrograph is an Andor Kymera 328i with protected-silver coated optics and a motorized entrance slit and motorized camera mirror for focus adjustments. The optics are aligned in a Czerny-Turner configuration with a fold mirror just after the side-port entrance and all reflective optics. The Kymera 328i grating turret holds 4 Richardson ruled gratings with protected silver coatings (Table \ref{tab:gratings}) to provide a range of wavelength and resolution coverage. Spectra are captured by an Andor iDUS 420 CCD, a BEX2-DD model with 256x1024, 25 \microm \ pixels. The CCD is back-illuminated, deep-depletion, with a dual AR coating cooled by thermo-electric cooling to -90C. Quantum efficiency is above 50\% from $\sim$350 -- 950 nm. 

\begin{table}[ht]
\begin{center}
\begin{tabular}{|c|c|c|c|}
\hline
Grating Ruling & Resolution (nm) & Blaze Angle (nm) & Wavelength Coverage (nm) \\ \hline
150 l/mm       & 1.46            & 800              & $\sim$500 -- 1100        \\ \hline
600 l/mm       & 0.35            & 500              & $\sim$350 -- 700         \\ \hline
600 l/mm       & 0.35            & 1000             & $\sim$850 -- 1400        \\ \hline
1200 l/mm      & 0.164           & 500              & $\sim$400 -- 850         \\ \hline
\end{tabular}
\end{center}
\caption{Kymera 328i Grating properties \label{tab:gratings}}
\end{table}

For simultaneous imaging, target acquisition, and tracking, PEAS uses an SBIG STF-8300 imager, with 3326 x 2504 pixels at 5.4 \microm. A filter slot holds 31 mm Bessel UBVRI filters from Chroma. The field of view of the camera is $\sim$ 17\arcmin x 13\arcmin. The imager works seamlessly with the Planewave Telescope software to communicate tracking offset information.

\subsection{Mechanical Design and Electronics}
\begin{figure} [h!]
  \begin{center}
  \includegraphics[height=6.8cm]{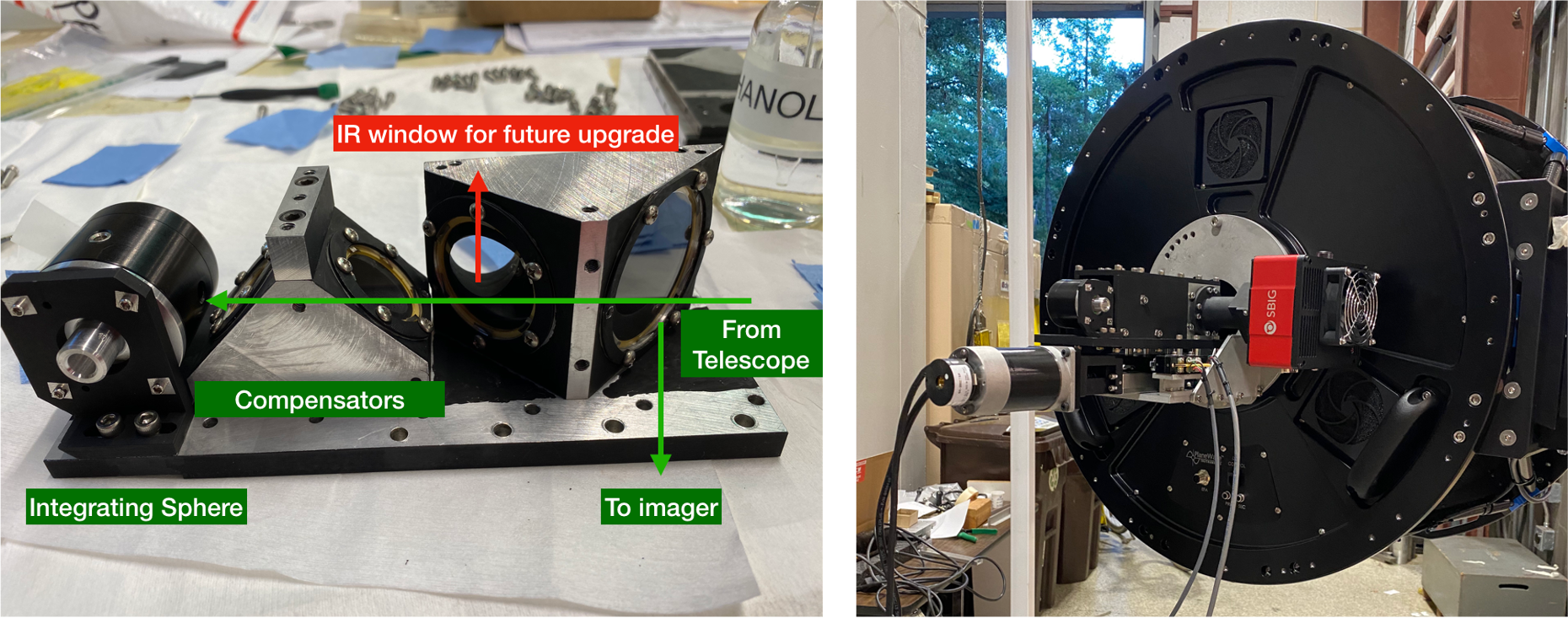}
  \end{center}
  \caption 
  { \label{fig:optics} 
\textit{Left}: Beamsplitter assembly. Sapphire windows are mounted into V blocks of aluminum and mounted onto an aluminum base plate. The windows are held in place by thin titanium laser-cut springs. The general light paths are shown in green to the imager and integrating sphere, and the optional window for a future IR upgrade has been accounted for in the design. \textit{Right}: Once assembled, the optics sit atop a THK linear stage run by a galil motor and motion controller. The whole assembly attaches to a flange that matches the back plate of the Planewave telescope.}
  \end{figure} 

The beamsplitter assembly holds the windows, integrating sphere, and imager and bolts to the back of the telescope with an aluminum flange. The Sapphire windows are mounted into aluminum V-shaped blocks cut at 45 degrees. The windows sit inside cutouts and are held in place axially by thin titanium laser-cut springs that screw into the V-block with a retainer ring. Thin plastic rings protect the windows from the aluminum surface and the titanium springs. The V-blocks are mounted onto an aluminum plate and covered on all sides with aluminum plate painted with Krylon Interior/Exterior black spray paint. The optical tolerances are loose enough to allow for a ``bolt and go" approach to optical alignment of the windows. Once the windows are fully mounted in the assembly the alignment is ready for observations and does not need adjustments to align the optics. The integrating sphere is held by an L shaped bracket that mounts to the fiber ferrule exit port. The bracket is slotted to allow for focus adjustments. The optical assembly is shown in Figure~\ref{fig:optics} at left. 

The four-sided aluminum structure that holds the beamsplitters attaches to an aluminum tube and with a flange to secure the SBIG imager in place. The tube has a slot for 31 mm optical filters. The optics plate sits atop a base plate on the THK linear stage, which acts as a focus mechanism (Figure~\ref{fig:optics}, right). The linear stage is driven by a galil motor and controlled by a galil DMC 30012 motion controller. Primary limit switches use Hall-effect transistors to locate Home, Forward, and Reverse limits and the secondary limit switch is a mechanical rocker switch. The switches and switch harness are wired similarly to the Lick Automated Planet Finder linear stages [\citenum{vogt2014}]. 
  
  \begin{figure} [h!]
  \begin{center}
  \includegraphics[width=15cm]{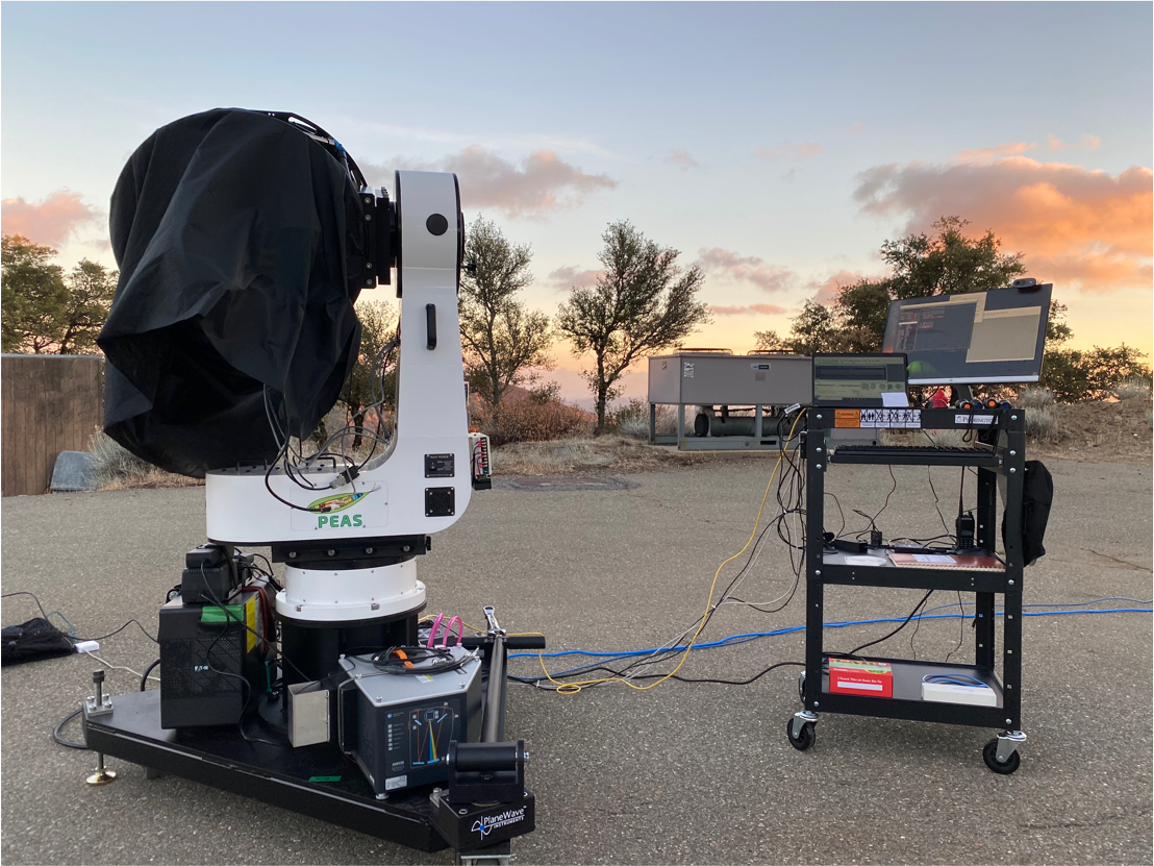}
  \end{center}
  \caption 
  { \label{fig:lick} 
PEAS at Lick Observatory just before achieving first-light on 25 November 2020. The telescope and rolling pier mount are on the left, looking toward the southern skies, just east of the Shane 3-m dome on Mt. Hamilton. The computer cart at right holds the Windows machine that runs the software for the Planewave, SBIG, and Andor.}
  \end{figure} 
  
The full instrument assembled at Lick observatory and ready for first-light observations is shown in Figure~\ref{fig:lick}. The telescope back-end is baffled with black vinyl from Thorlabs, which allows the cabling from the motor and switches, SBIG camera, and the optical fiber to feed through easily. The electronic cabling wraps through the L500 mount to an opening in the rolling pier mount so that cables do not get tangled or over-wrapped. A separate computer cart holds the Windows machine that runs software for the SBIG imager, Andor spectrograph, and Planewave telescope. 100-ft ethernet and power cords run from inside the Shane Dome to power the telescope and allow communication through the Mt. Hamilton network. Table~\ref{tab:components} contains a detailed list of PEAS components. 

\begin{table}[h!]
\begin{center}
\begin{tabular}{|p{0.18\linewidth}|p{0.18\linewidth}|p{0.28\linewidth}|p{0.28\linewidth}|}
\hline
\textbf{Component} & \textbf{Manufacturer} & \textbf{Specifications} & \textbf{Purpose} \\ \hline
Telescope & Planewave & RC 20" & Observe Planets \\ \hline
Sapphire Windows & Knight Optical & (50, 40, 30, 25) mm diameter x 2 mm thickness, random cut & Beamsplitters\\ \hline
Integrating Sphere & Spectral Products & 1" sphere & Disk-integrate light \\ \hline
Fiber bundle & MaNGA/SDSS & 120 \microm \ core 32 fiber pseudo-slit & Feed light from integrating sphere to spectrograph \\ \hline
Spectrograph & Andor & Kymera 328i & Disperse planet light \\ \hline
Spectrograph CCD & Andor & iDUS 420 BEX2-DD & Record spectra \\ \hline
Imager & SBIG & STF8300 & Simultaneous imaging, pointing, tracking \\ \hline
Optical Filters & Chroma & UBVRI 1.25" Bessel & Filters for optical imager \\ \hline
Linear Stage & THK & SKR26 & Focusing \\ \hline
Motor & Galil & BLM-N23-50-1000-B Brushless motor & Drives linear stage \\ \hline
Motion Controller & Galil & DMC30012 & Controls Motor \\ \hline
UPS & Eaton & 9SX-3000 2700 W & Online UPS, backup battery \\ \hline
Windows Laptop & Lenovo & Thinkpad X390 + Windows 10 & Run software for Andor, SBIG, Planewave \\ \hline
Linux machine & Intel & NUC i7 with CentOS 8 & Run linear stage, communications with Mt. Hamilton network \\ \hline
Router & Ubiquity & EdgeRouter x & Communications \\ \hline
Switch & Netgear & 8-port & Communications \\ \hline
Andor CCD Chiller & Koolance & EXT-440CU Chiller & Liquid coolant chiller of the spectrograph CCD \\ \hline
Cover & Ullman Sails & Custom cover made from Sail material & Cover PEAS telescope and instrument while not in use \\ \hline
Baffle & Thorlabs & Black Vinyl Cloth & Baffle the optics \\ \hline
Rat Repellent & Angveirt & Ultrasonic Rodent Repeller & Repel rats and mice \\ \hline

\end{tabular}
\end{center}
\caption{PEAS components and specifications \label{tab:components}}
\end{table}

\section{Initial Results}
\label{sec:firstlight}

  \begin{figure} [h!]
  \begin{center}
  \includegraphics[width=13cm]{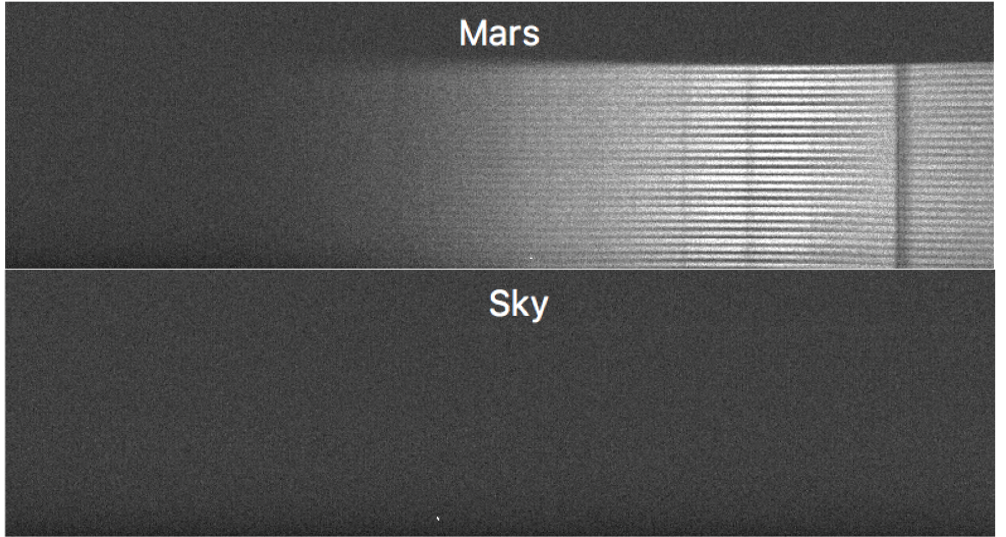}
  \end{center}
  \caption 
  { \label{fig:firstlight} 
PEAS first light data on Mars (60 second exposure) and an exposure of equal length of the Sky taken on 25 November 2020. Data were taken with the lowest resolution mode (R$\sim$500) covering $\sim$300 -- 800 nm. }
  \end{figure} 

PEAS was assembled in the shops at UC Observatories over Summer and Fall 2020 and was moved to Lick Observatory in mid-November, 2020. PEAS achieved First Light on 25 November, 2020. Our first observations were of Mars (Figure~\ref{fig:firstlight}) using the lowest resolution grating (R$\sim$500) covering $\sim$300--800 nm. An equal-length exposure of the sky is shown for reference. Over Winter 2020-2021, PEAS will be commissioned, and then begin regular science operations in early 2021. The first science results will be an atlas of Solar System planet spectra observed as if they are point-source exoplanets. These data will be compared to ``ground truth" observations of Solar System planets from existing data from fly-by's, rovers, landers, probes, and ground- and space-based observations. Additionally, the data will be used to compare and update existing planetary atmospheric models and will inform the next generation of exoplanet instrument design.

\acknowledgments 
 
PEAS is supported by the Heising Simons Foundation (Grant \# 2019-1681). ECM acknowledges support from NSF Astronomy and Astrophysics Postdoctoral Fellowship under award AST-1801978.

\bibliography{peas} 
\bibliographystyle{spiebib} 

\end{document}